\documentclass{pasj00}
\draft

\begin{document}
\SetRunningHead{Deguchi et al.}{HOCO$^+$ toward the Galactic Center}
\Received{2006/07/10}%{yyyy/mm/dd}
\Accepted{; Ver 1.1 Sept 18, 2006}%{yyyy/mm/dd}

\title{HOCO$^+$ toward the Galactic Center}

%%% begin:list of authors
\author{Shuji \textsc{Deguchi},$^{1}$ Atsushi \textsc{Miyazaki},$^{1,2}$} 
\and 
\author{Young Chol \textsc{Minh}$^{3,4}$}
%%%%%%%%%%%%%%%%%%% following is the note of the version %%%%
%\author{\\(will submitted to PASJ letter --- Version: 2001/07/15)\\}

\affil{$^{1}$ Nobeyama Radio Observatory, National Astronomical Observatory,\\
              Minamimaki, Minamisaku, Nagano 384-1305}    
\affil{$^{2}$ Shanghai Astronomical Observatory, Chinese Academy of Sciences \\
              80 Nandan Road, Shanghai, 200030, P.R. China}
\affil{$^{3}$ Korea Astronomy and Space Science Institute, \\
              61-1 Hwaam, Yuseong-ku, Daejeon 305-348, Korea}
\affil{$^{4}$ Academy Sinica, Institute of Astronomy and Astrophysics,
              P.O. Box 23-141, Taipei 106, Taiwan}   
%\affil{$^{10}$ Department of Astronomy, University of Illinois at Urbana-Champaign\\
%1002 W. Green St., Urbana, IL 61801-3074, U.S.A.}

% \author{\\(PASJ  April 28 issue in press--- Version: 2004/02/27)\\}

%\affil{C-Address of Institute}\email{ccccc@xxx.xxx.xx.xx}
%%% end:list of authors

%%% Please use the following style in case that sorting by 
%%% affilation is impossible. 
%
% \author{%
%   D-Firstname \textsc{D-Familyname}\altaffilmark{1}
%   E-Firstname \textsc{E-Familyname}\altaffilmark{1,2}
%   and
%   F-Firstname \textsc{F-Familyname}\altaffilmark{2}}
% \altaffiltext{1}{Address of Institute}
% \email{ddddd@xxx.xxx.xx.xx}
% \email{eeeee@xxx.xxx.xx.xx}
% \altaffiltext{2}{Address of Institute}

%% `\KeyWords{}' always has to be placed before `\maketitle'.
\KeyWords{Galaxy:  center --- ISM: molecules --- line: identification ---
molecular data} %Do NOT move this preamble from here!

\maketitle

\begin{abstract}
We have identified a weak thermal line U42.767,  
which has been detected only in the directions toward Sgr A and Sgr B2,
as the HOCO$^+$ $2_{02}$--$1_{01}$ transition.
Because of the proximity of this line to the SiO maser line at 42.821 GHz ($J=1$--0 $v=$2),
it was observable simultaneously in the $\sim 43$ GHz SiO maser searches at Nobeyama.
From the past data of SiO maser surveys of infrared objects in the Galactic center,
we created a map of emission distribution of HOCO$^+$ in the Sgr A molecular cloud
as well as maps of the $^{29}$SiO $J=1$--0 $v=0$ thermal emission and H53$\alpha$ emission.
The emission distribution of HOCO$^+$ was quite similar to
the distribution of $^{29}$SiO emission. It suggests that the enhancement of the HOCO$^+$ abundance
in the galactic center is induced by shock activities which release the CO$_2$ molecules frozen on grains
into gases. 
 \end{abstract}

% Figure 1   spectra
% Figure 2a-c  integrated spectra
% Figure 3a-c  channel map.

\section{Introduction} 

In the course of the past SiO maser line observations in the 43 GHz band at Nobeyama, 
we noticed that several weak unidentified lines occasionally contaminated
the spectra toward stars in the Galactic center (\cite{deg02,deg04}). 
They are not listed in the catalog of known molecular lines compiled 
by \citet{lov04}.\footnote{For latest compilation available in 
{\it http://physics.nist.gov/PhysRefData/Micro/Html/contents.html}.} 
Among these, the unidentified line, U42.767, bothered us strongly 
because it appears near the SiO $J=1$--0 $v=2$ 
line at 42.821 GHz with a velocity separation of about 380 km s$^{-1}$,
making a false signal of extreme high-velocity maser emission toward the Galactic center.
The width of this unidentified line was quite broad compared with SiO maser lines,  
indicating that it is thermal emission from molecular clouds. 
On the high frequency side of the SiO line, the other weak lines of
$^{29}$SiO $J=1$--0 $v=0$ line ($\Delta V=-416$ km s$^{-1}$) and 
H53$\alpha$ ($\Delta V=-921$ km s$^{-1}$) also contaminate the SiO maser spectra.  
A comparison of emission intensities between U42.767, 
$^{29}$SiO, and H53$\alpha$ suggested that U42.767 appears only in a relatively dense region 
of molecular clouds in the Galactic center.
So far, we have surveyed toward many AGB stars in the Galactic bulge and disk, 
and several molecular clouds such as Ori A, TMC1, and W51 in these lines, but
we did not detect U42.767 except in the Galactic center region (see \cite{izu95,miy01,kai04}).
   
By checking the JPL catalog of Molecular Spectroscopy (\cite{pic98}),\footnote{
The latest version is available at {\it http://spec.jpl.nasa.gov/}}
 we found that one of the best candidates
for U42.767 is the HOCO$^+$ $2_{02}$--$1_{01}$ transition (42766.2 MHz).
The other rotational lines of this molecule have been detected
only in the molecular clouds toward the Galactic center: Sgr A and Sgr B2
(\cite{tha81,min88}).     
 
In 2001--2003, we made a large SiO maser survey of about 400 stars within 15$'$ of Sgr A*
(\cite{ima02,deg04}). This survey took the data of the above three lines in one spectrometer
with a band width of 250 MHz, together with the SiO $J=1$--0 $v=2$ maser line.
In this paper, we describe the result of mapping in these three lines
toward the Sgr A molecular cloud. The emission distribution of U42.767 was quite similar to
the distribution of the HOCO$^+$ $4_{04}$--$3_{03}$ line at 85.531 GHz
obtained before by \citet{min91}. Based on all of these results, 
we conclude that U42.767 is the HOCO$^+$ $2_{02}$--$1_{01}$ transition.
  
%              45010  HOCO$^+$        
%   42766.1975  0.0107 -4.3892 3    0.7133  5  45010 303 2 0 2       1 0 1       

%             45010  HOCO$^+$        
%   42598.1928  0.0090 -4.5716 3   26.6663  5  45010 303 2 1 2       1 1 1       
%   42766.1975  0.0107 -4.3892 3    0.7133  5  45010 303 2 0 2       1 0 1       
%   42926.7971  0.0181 -4.5650 3   26.6718  5  45010 303 2 1 1       1 1 0       

%\newpage

\section{Observational data and results}
We used the spectral data taken by an acousto-optical spectrometer
with 250 MHz band width (AOS-W) in the course of an SiO maser survey
of the large amplitude variables in the Galactic center (\cite{ima02}).
In these observations, data at the positions of about 400 stars toward the Galactic center
(within 15$'$ from Sgr A*) were obtained during 2001--2003 
with the 45-m radio telescope at Nobeyama.  
The effective frequency resolution of
the AOS-W spectrometer is 250 kHz (1.75 km s$^{-1}$). 
The overall system temperature was between 200 and 300 K,
depending on weather condition. The half-power telescope beam width
(HPBW) was about 40$''$ and the beam efficiency at 43 GHz was 0.77. 
Typical rms noise level in the spectra was about 0.03 K.
Further details of SiO maser observations using the NRO 45-m telescope have been
described elsewhere (\cite{deg04}), and are not repeated here.
The AOS-W spectra occasionally exhibited considerable baseline distortions.
We applied a parabolic baseline removal to the individual spectra and
extracted intensities of the HOCO$^+$ $2_{02}$--$1_{01}$, 
$^{29}$SiO $J=1$--0 $v=0$, and H53$\alpha$ transitions
from the spectra. The rest frequencies used are given in table 1.     
Because the lines were weak and because standing-wave features (appearing occasionally)  
could not be removed completely,  we discarded a few bad data.  
The data points used in the map making are not uniformly distributed
for the three mapped lines: the number of points used for map making are
369, 370, and 347 for the HOCO$^+$ $2_{02}$--$1_{01}$, $^{29}$SiO $J=1$--0, and H53$\alpha$ lines. 
Because of non-uniform sampling of the observed points,  
we applied a 140$''$ Gaussian smoothing in the map making; 
the effective spatial resolution in the created map was lowered
in this procedure. 

Figure 1 shows the AOS-W spectra toward the SiO maser sources [GMC2001] 12--13 (=V5038 Sgr; see \cite{gla01} or
\cite{deg04})
and Sgr B2 MD5; the latter does not exhibit the SiO $J=1$--0 $v=2$ emission (see \cite{shi97}). 
Adjacent to the SiO $J=1$--0 $v=2$ maser line at 42.821 GHz, the 
HOCO$^+$ $2_{02}$--$1_{01}$, $^{29}$SiO $J=1$--0, and H53$\alpha$ lines are recognizable.
Figures 2a--2c show the integrated intensity maps of the above three lines toward the Galactic center.
%within a velocity range of $\pm 100$ km s$^{-1}$.
The  velocity channel maps (combined in every 20 km s$^{-1}$)  are also shown in figures 3a--3c.
Figure 2a indicates that the emission distribution is quite similar to 
the HOCO$^+$  4$_{04}$--3$_{03}$ emission shown in figure 1b of \cite{min91}, suggesting the validity of 
identification of U42.767 to this molecule; both transitions have triple intensity peaks at
$(l, b)=(0.11^{\circ}, -0.08^{\circ})$, $(0.01^{\circ}, -0.07^{\circ})$ and $(-0.13^{\circ}, -0.08^{\circ})$.

We also checked the more recent AOS-W data taken by the 2004--2005 SiO maser survey 
toward the $7^{\circ} \times 2^{\circ}$ area of the Galactic center (\cite{fuj06}). The surveyed area
covered the Sgr C molecular cloud (G395.43$-$00.09). However, no HOCO$^+$ emission
stronger than 0.10 K was detected, especially toward MSX objects near Sgr C:
G359.2871$-$00.2009, G359.3258+00.0040, G359.3819+00.0049,
and G359.5035$-$00.1073. The Sgr C cloud, as well as Sgr A and B clouds,  
is known as a 6.4 keV X-ray reflection cloud (\cite{mur03}).

%%%%%%%%%%%%%%%%%%%%%%%   Section 3. %%%%%%%%%%%%%%%%%%%%%%%%%%%%%%%%%%

\section{Discussion}

\subsection{Chemistry and excitation of HOCO$^+$}

The HOCO$^+$ molecule (protonated carbon dioxide) is thought to form via a standard ion-molecule reaction, 
the transfer of a proton from H$_3^+$ to CO$_2$ (\cite{her77,tur99}). 
Interestingly, the extensive survey of the HOCO$^+$ toward various Galactic objects revealed that 
this molecule is exclusively abundant only toward our Galactic center region (Minh et al.\ 1988; 1991).
Since HOCO$^+$ is the protonated ion of CO$_2$, the enhancement of the HOCO$^+$ abundance may suggest 
an increase of the gas phase CO$_2$ abundance.
Though CO$_2$ may be an important reservoir of interstellar carbon and oxygen, 
the lack of an electric dipole moment of this symmetric molecule 
made gas phase detections difficult.
Observations of the 15 $\mu$m band of CO$_2$ from space %(\cite{jus98})
indicate that the observed gas-phase CO$_2$ abundances 
are typically $\sim 2 \times 10^{-7}$ 
toward various star-forming regions, which is up to two orders 
of magnitude lower than those of solid CO$_2$ in the same regions (\cite{van04}).

The chemistry connecting CO$_2$ with HOCO$^+$ is not completely clear. However, 
if HOCO$^+$ forms by the reaction 
with the gas phase CO$_2$, the large enhancement of HOCO$^+$ in our Galactic center 
indicates that the special environment of the Galactic center is necessary to enhance 
the CO$_2$ abundance exclusively in this region.

Excitation of the HOCO$^+$ rotational levels was discussed extensively 
in \citet{min91}. 
% The abundance of HOCO$^+$ toward Sgr A was obtained  to be $8 \times 10^{-9}$ per H$_2$. 
The observed line intensity ratio of $4_{04}$--$3_{03}$ to $2_{02}$--$1_{01}$ is
% $T_R^*(4_{04}-3_{03})/T_R^*(2_{02}-1_{01})= 
1.56 % where $T_R^*$ is the brightness temparature of the line,  %%% (Ratio = 5.9/0.73 / 4/0.77)
at the HOCO$^+$ peak position [$(l,b)=(0.11^{\circ}, -0.08^{\circ})$], which 
gives a rotational temperature of 7.0 K (assuming the lines to be optically thin),
and the HOCO$^+$ column density of $\sim 4.2 \times 10^{13}$ cm$^{-2}$. 
[At this position the excitation of HOCO$^+$ may not be affected much by the background infrared radiation 
field which can enhance the {\it b}-type transitions of this slightly asymmetric top, 
as was discussed by Minh et al.\ (1991) in their excitation model calculation.]
In order to derive the fractional abundance of HOCO$^+$, we use the total H$_2$ column density, 
$\sim 6 \times 10^{22}$ cm$^{-2}$, derived by Minh et al.\ (2005) at this position. Then, 
we obtain the  fractional abundance of HOCO$^+$, $\sim 7 \times 10^{-10}$, relative to H$_2$;
this value is about an order of magnitude less than the one derived 
using an excitation model previously by Minh et al.\ (1991). 
The fractional abundance of $\sim 10^{-9}$ is still much larger, 
by more than an order of magnitude,
than the value expected from the gas phase reactions (see discussions in Minh et al.\ 1988).
However, if we apply the large H$_2$ column density, $\sim 6 \times 10^{23}$ cm$^{-2}$, 
suggested by Handa et al.\ (2006) at G 0.11$-$0.11,
the fractional abundance of HOCO$^+$ will be lowered as well.
The abundances for this cloud, which shows an HCO$^+$ enhancement (Minh et al.\ 2005),
will be further discussed in the following section.
We think that the HOCO$^+$ peaks in the Sgr A region certainly trace a specific physical
condition which can increase the HOCO$^+$ formation,  such as the shock induced grain evaporation.
HOCO$^+$ also shows, unlike most other molecules, a unique distribution in Sgr B2, 
which strongly suggests that the formation of this molecule needs a specific condition existing
exclusively in our Galactic center \citep{min98}.
%

%%%%%%%%%%%%%%%%%%%%%%%%%%%%%%%%%% subsection molecular clouds %%%%%%%%%%%%%%%%%%%%%%%%%%%%%%%%%%%%%%%%%%%%
\subsection{the Galactic center molecular clouds}
The region within $\sim 20'$ from the Galactic center (Sgr A*) is
one of the best investigated areas in various molecular/atomic lines and continuum 
(for example, see \cite{oka98,tsu99}).
There are several dense molecular clouds including Sgr A East 
(\cite{gus81,arm85}) and G0.11$-$0.11 (\cite{han06});
the latter is known to have an extremely high H$_2$ column density (\cite{han06}). 
However, the molecular clouds in this area are known to be clumpy; 
\citet{miy00} analyzed the mass spectra of the clumps using CS emission 
in this area.   

% \subsection{Surface Density}

Though the general morphology of HOCO$^+$ emission in figure 2a is similar 
to the previous results obtained from the $4_{04}-3_{03}$ transition
(figure 1b of \cite{min91}), 
it shows slightly higher contrast, 
partially due to different position sampling and applied smoothing 
(''random'' sampling in this paper),
and partially due to the excitation of the two observed transitions
(the excitation energies from the ground state to the $2_{02}$ and $4_{04}$ levels 
are about 3 and 10 K, respectively). 
The rightmost bright clump corresponds to the gas condensation M$-0.13-0.08$ 
(indicated by the rightmost filled square in figure 2a), 
which is known as the +20 km s$^{-1}$ cloud.
\citet{min05} argued that the cloud M$-0.13-0.08$ consists of two different velocity components, 
the 5 km s$^{-1}$ and 25 km s$^{-1}$ components, 
based on the chemical differences and steep velocity gradient.
The 25 km s$^{-1}$ cloud is located near to the expanding SNR Sgr A East and is
thought to be interacting with it directly 
\citep{oku91,ho91,coi00}.
The HOCO$^+$ $4_{04}-3_{03}$ emission peaks toward the center of the +25 km s$^{-1}$ cloud, 
but the $2_{02}-1_{01}$ emission peaks at the position $2^\prime$--$3^\prime$ lower in $b$.
However, this is probably an artifact due to undersampling of the observed positions
at the lower part of this cloud. The line intensity ratio
of the $2_{02}$--$1_{01}$ to $4_{04}-3_{03}$ transition suggests
a considerably high excitation temperature ($\gtrsim 200$ K) at the center of M$-0.13-0.08$.
If the +25 km s$^{-1}$ component is interacting with Sgr A East toward the Galactic center,
the high excitation temperature may indicate that the HOCO$^+$ is actually tracing 
the interacting region.
 
Another HOCO$^+$ emission peak is located near the cloud M$-0.02-0.07$
(indicated by the central filled square in figure 2a),
which is known as the +50 km s$^{-1}$ cloud. Both
M$-0.02-0.07$ (the +50 km s$^{-1}$ cloud) and M$-0.13-0.08$ (the +20 km s$^{-1}$ cloud) 
are major gas condensations in the Sgr A region, having total masses of the order of $\sim 10^5$ M$_\odot$
(e. g., \cite{gus81,arm85}).
M$-0.02-0.07$ also has been thought to be associated with the observed ``molecular ridge" compressed
by the expanding Sgr A East cloud, which lies just outside Sgr A East (\cite{ho85,gen90,ser92}).
The elongated shape of the HOCO$^+$ emission in this region well resembles 
the distribution of the molecular ridge, but it does not agree with the general +50 km s$^{-1}$ cloud
morphology.
If the 50 km s$^{-1}$ cloud (M$-0.02-0.07$) is chemically quiescent 
and located far from the Galactic center compared to other condensations [as \citet{min05} argued],  
this HOCO$^+$ elongation indicates that the abundance of this molecule also increases 
in the interacting region, like the +25 km s$^{-1}$ component of M$-0.13-0.08$.
This cloud also coincides with the enhanced H$_2$ emission region, 
which is a region being shocked by interaction between the Sgr A East cloud and the ambient gas; 
it is likely to be assiciated with the +50 km s$^{-1}$ cloud (\cite{lee06}).

The brightest HOCO$^+$ peak in the observed field was found at $(7', -5')$, 
which is close to the M$0.11-0.08$ (indicated by the leftmost filled square in figure 2a).
This region has been considered to have a relatively smaller column density 
among the known condensations in this field (see \cite{arm85}). 
Interestingly, the HCO$^+$ emission exhibits the strongest peak $2'$ south of the HOCO$^+$ peak
near M$0.07-0.08$ \citep{min05}.
The large enhancement of HCO$^+$ compared to $^{13}$CO at this position 
strongly suggests that the formation of HCO$^+$  in this region does not proceed  
with CO, but with other shock related reactions involving the molecules such as C$^+$ or OH. 
\citet{han06} showed a shell-like morphology of the SiO $J=1$--0 thermal emission in this region
(which they assigned as G$0.11-0.11$).
Their two peaks, A and B, 
which were not resolved in figures 2a and 2b due to the applied smoothing,
coincide with our HOCO$^+$ and $^{29}$SiO $J=1$--0 peaks in position. 
Furthermore, their ridge (named E), which gives a strong peak in H$^{13}$CO$^+$ emission,
was located 2$'$ south of the HOCO$^+$ peak, where \citet{han06} found the peak of the H$_2$ column density. 

This region contains a lot of thermal and non-thermal features, 
indicating that shocks and high-energy photons 
play significant roles 
in characterizing this region (\cite{fig02,yus03}). The 6.4 keV neutral Fe line
was also detected in this region \citep{mur03}. 
Thus the bright HOCO$^+$ emissions are all related with the regions 
which are thought to be affected by the strong Galactic center activities.
This fact is a clue to understand the HOCO$^+$ emission in the Galactic-center molecular clouds. 
It also indicates that HOCO$^+$ may be a good tracer to probe the activities of the Galactic center.

Figure 2b shows the integrated intensity map of the $^{29}$SiO $J=1$--0 emission. 
It also exhibits triple peaks similar to HOCO$^+$ in figure 2a, 
which can be compared with the SiO $J=1$--0 emission map \citep{mar97}.
For the component associated with M $0.11-0.08$, 
our $^{29}$SiO $J=1$--0 (and HOCO$^+$) distribution exhibits a good correlation 
with the SiO $J=1$--0 thermal emission (\cite{han06}) in general; 
figures 2a and 2b indicate stronger peaks at the positions A and B
(though they were not resolved in our maps as mentioned),
suggesting the SiO $J=1$--0 emission is optically thick [as found by \citet{han06}]. 

It is well known that silicon, like the other heavy elements, is locked up 
in the refractory materials and shows a strong depletion in the gas phase in inactive dense clouds. 
The fractional abundance of the gas phase SiO relative to H$_2$ 
is observed to be less than $10^{-7}$ (\cite{ziu89}).
Enhancement of SiO in the gas phase is thought to be related with shock waves, 
which are expected to erode and to partly destroy the dust through sputtering 
and grain-grain collisions (e.g., \cite{pin97}).
The coincidence of the $^{29}$SiO and the HOCO$^+$ emission regions again strongly suggests 
that the HOCO$^+$ enhancement is related with the shocks or Galactic center activities.

Figure 2c shows the H53$\alpha$ recombination line distribution. The emission peaks are located 
outside of the dense gas clouds. The strongest peak roughly coincides with HII region G0.10+0.02, 
observed by the 8.3 GHz continuum and H92$\alpha$ (\cite{lan01}). It also
coincides with the shocked diffuse cloud traced by the HCO$^+$ (\cite{min05}). In addition the well known arcs 
of the Galactic center are also located in this region. Therefore this recombination line 
traces diffuse shocked gas, which may result in the complicated ionized gas distribution.

\section{Conclusion}
We assigned the unidentified line U42.767 to the $2_{02}$--$1_{01}$ transition of HOCO$^+$.
The emission distribution in the Galactic center is quite similar to those
found previously in the higher rotational transitions of this molecule.
The past SiO maser surveys at $\sim 43$ GHz made at Nobeyama, which included the
above transition in the same spectrometer, confirmed the rare existence of
this ion except in Galactic center molecular clouds. It strongly suggests that 
CO$_2$ molecules are tightly frozen on dust grains in most of the Galactic clouds, 
but they are released from grains 
with other molecules such as SiO by shock activities in the Galactic center.  

\

Authors thank Prof. William Irvine, University of Massachusetts, 
for reading the manuscripts, and
Dr. Toshihiro Handa, University of Tokyo, 
and Dr. Shuro Takano, Nobeyama Radio Observatory, 
for helpful comments.
They also thank the SiO maser survey team 
for allowing use of the data of the 2001--2003 and 2004--2005 SiO maser surveys. 
One of the authors (A. M.) thanks the National Astronomical Observatory for
providing him with a visiting fellowship. 

%%%%%%%%%%  Appendix %%%%%%%%%%%%%
%\section*{Appendix. Indivisual Objects}
%\begin{itemize}

%\end{itemize}
%%%%%%%%%%%%%%%%%%%%%%%%%%%%%%%%%%%%%%%

%%%

%%%%%%%%%%%%%%%%%%%Tables %%%%%%%%%%
\clearpage
\begin{longtable}{lcll}
  \caption{Rest frequencies adopted.} 
\hline  \hline
species & transition & frequency & reference \\
      &            &  (MHz)     &   \\
\hline
\endfirsthead
\hline \hline
specy & transition & Frequency & A  \\
         &            &  (MHz)     & (s$^{-1}$) \\
\hline
\endhead
\hline
\endfoot
\hline
%\multicolumn{4}{l}{$^{\sharp}$ : .......}@\\
%\multicolumn{4}{l}{$^{\flat}$ : ........}@\\
\endlastfoot
HOCO$^+$ & $J_{K_a,K_b}=2_{02}$--$1_{01}$ & 42766.198 & \cite{pic98} \\
SiO  & $J=1$--0 $v=2$  &42820.582 & \cite{lov04} \\ 
$^{29}$SiO  & $J=1$--0 $v=0$  & 42879.922 & \cite{lov04}\\ 
H & 53$\alpha$ & 42951.97 & \cite{lil68} \\
He & 53$\alpha$ & 42969.48 & \cite{lil68} \\
\end{longtable}

%%%%%%%%%%%%%%%%%%%%%%%%%%%%%% End of this paper %%%%%%%%%%%

% \setcounter{page}{0}
%%%%%%%%%%%%%%%%%%%%%%% figure 1 %%%%%%%%%%%%%%%%%%%%%%%%

\begin{figure}
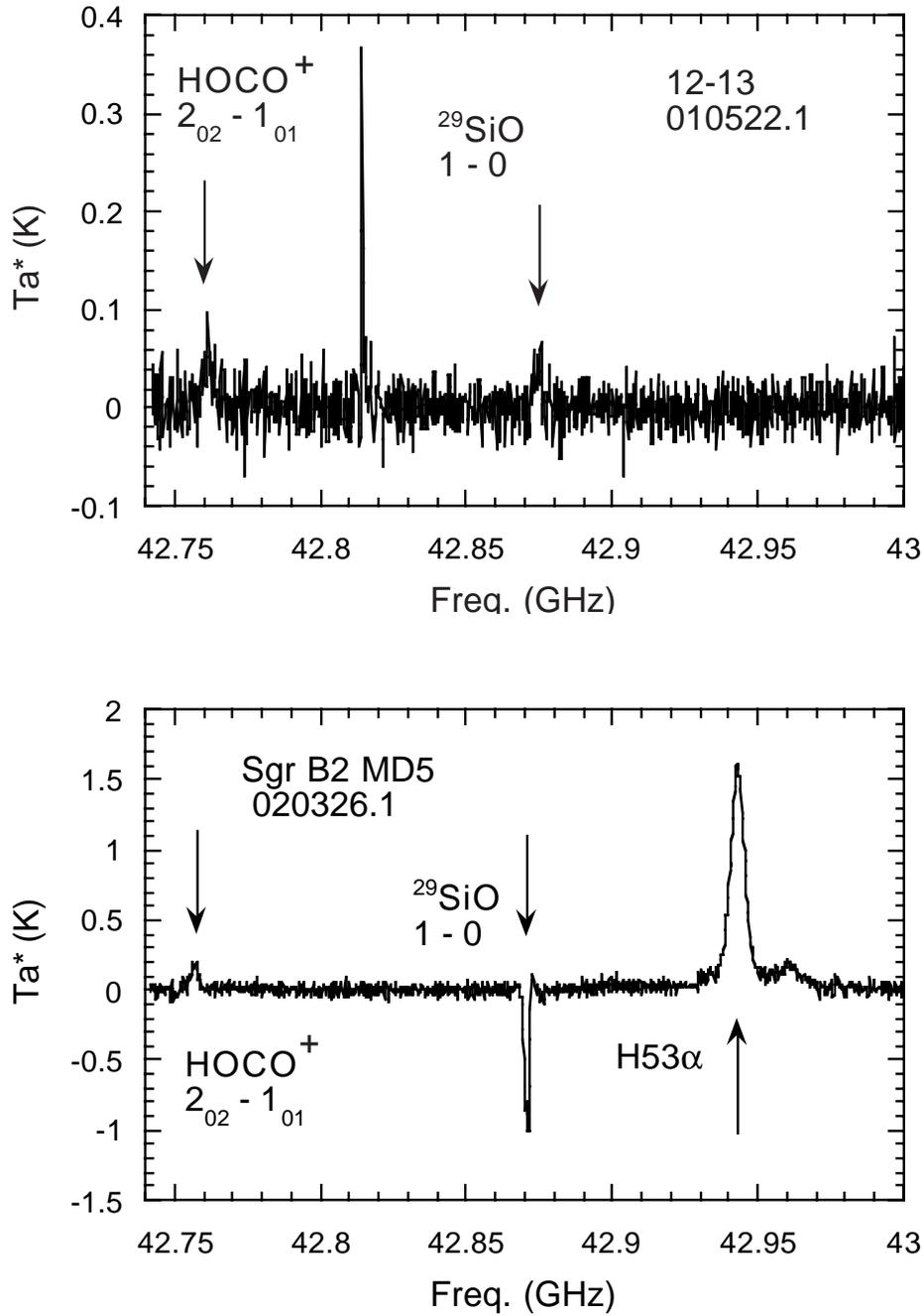

  \begin{center}
    \FigureFile(140mm,180mm){fig1.eps}
  \end{center}
  \caption{AOS-W Spectra between 42.74 and 43.0 GHz toward [GMC2001] 12--13 
  (upper panel) and toward Sgr B2 MD5 (lower panel). The observational date was shown 
  in yymmdd.d format under the object name. 
  The sharp strong line at 42.82 GHz in the upper panel 
  is the SiO $J=1$--0 $v=2$ maser line from [GMC2001] 12--13. On the right of H53$\alpha$, a weak enhancement
  due to He53$\alpha$ is recognizable in the lower panel. 
}\label{fig: fig1}
\end{figure}
%%%%%%%%%%%%%%%%%%%%%% figure 2a %%%%%%%%%%%%%%%%%%%%%%%%

\begin{figure}
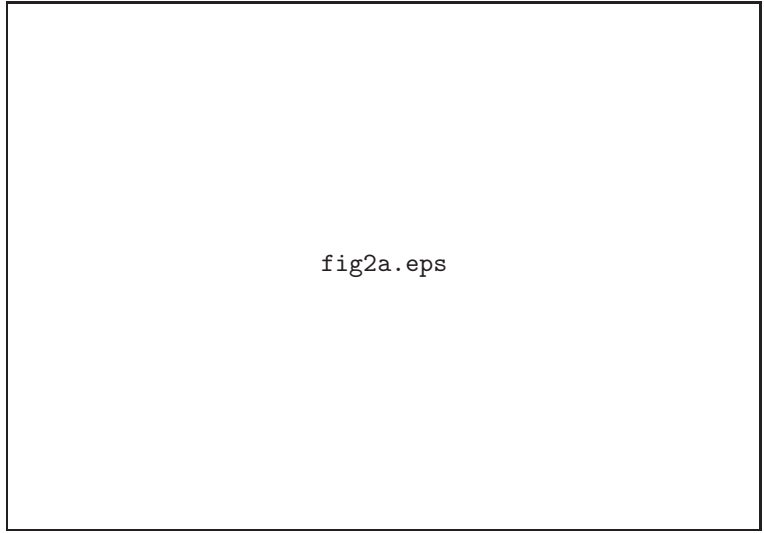

  \begin{center}
    \FigureFile(100mm,70mm){fig2a.eps}
    %%% \FigureFile(width,height){filename}
  \end{center}
  \caption{a. Integrated intensity map of the HOCO$^+$ $2_{02}$--$1_{01}$ emission toward the Galactic center. 
  The observed positions are marked by crosses.
  The coutours are drawn by every 0.3  K km s$^{-1}$ to the peak flux of 4 K km s$^{-1}$.   
  The outer lines near corners indicate the outer boundary of the sampled positions. 
  The filled squares indicate the positions of M0.11$-$0.08, M$-0.02-0.07$, and M$-0.13-0.08$,
  from the left to the right.
}\label{fig: fig2a}
\end{figure}
%%%%%%%%%%%%%%%%%%%%%%% figure 2b %%%%%%%%%%%%%%%%%%%%%%%%
\setcounter{figure}{1}
\setcounter{figure}{1}
\begin{figure}
  \begin{center}
    \FigureFile(100mm,70mm){fig2b.eps}
    %%% \FigureFile(width,height){filename}
  \end{center}
  \caption{b. Integrated intensity map of the $^{29}$SiO  $J=1$--0 $v=0$ emission toward the Galactic center. 
  The observed positions are marked by crosses.
  The coutours are drawn by every 0.3  K km s$^{-1}$ to the peak flux of 2.8 K km s$^{-1}$. 
  The filled squares indicate the positions of M0.11$-$0.08, M$-0.02-0.07$, and M$-0.13-0.08$,
  from the left to the right.
}\label{fig: fig2b}
\end{figure}
%%%%%%%%%%%%%%%%%%%%%%% figure 2c %%%%%%%%%%%%%%%%%%%%%%%%
\setcounter{figure}{1}
\begin{figure}
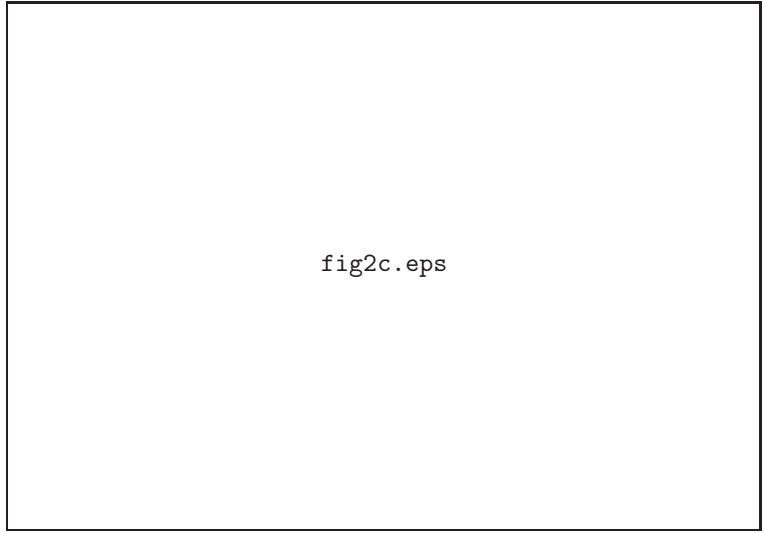

  \begin{center}
    \FigureFile(100mm,70mm){fig2c.eps}
    %%% \FigureFile(width,height){filename}
  \end{center}
  \caption{c. Integrated intensity map of H53$\alpha$ toward the Galactic center. 
  The observed positions are marked by crosses.
  The coutours are drawn by every 0.3 K km s$^{-1}$ to the peak flux of 3.5 K km s$^{-1}$. 
  The filled squares indicate the positions of M0.11$-$0.08, M$-0.02-0.07$, and M$-0.13-0.08$,
  from the left to the right.
}\label{fig: fig2c}
\end{figure}
%%%%%%%%%%%%%%%%%%%%%%% figure 3a %%%%%%%%%%%%%%%%%%%%%%%%
\setcounter{figure}{2}
\begin{figure}
  \begin{center}
    \FigureFile(170mm,200mm){fig3a.eps}
    %%% \FigureFile(width,height){filename}
  \end{center}
  \caption{a. Velocity-channel map of the HOCO$^+$ $J{KK}=2_{02}$--$1_{01}$ emission toward the Galactic center. 
 The coutours are drawn by every 0.01 K to the peak temperature of 0.101 K.
}\label{fig: fig3a}
\end{figure}
%%%%%%%%%%%%%%%%%%%%%%% figure 3b %%%%%%%%%%%%%%%%%%%%%%%%
\setcounter{figure}{2}
\begin{figure}
  \begin{center}
    \FigureFile(170mm,200mm){fig3b.eps}
    %%% \FigureFile(width,height){filename}
  \end{center}
  \caption{b. Velocity-channel map of the $^{29}$SiO  $J=1$--0 $v=0$ emission toward the Galactic center.
  The coutours are drawn by every 0.01 K to the peak temperature of 0.084 K.
}\label{fig: fig3b}
\end{figure}
%%%%%%%%%%%%%%%%%%%%%%% figure 3c %%%%%%%%%%%%%%%%%%%%%%%%
\setcounter{figure}{2}
\begin{figure}
  \begin{center}
    \FigureFile(170mm,200mm){fig3c.eps}
    %%% \FigureFile(width,height){filename}
  \end{center}
  \caption{c. Velocity-channel map of H53$\alpha$ toward the Galactic center.
  The coutours are drawn by every 0.01 K to the peak temperature of 0.081 K.
}\label{fig: fig3c}
\end{figure}


\begin{thebibliography}{}
\bibitem[Armstrong et al.(1985)]{arm85} Armstrong, J. T., Barrett, A. H., 1985, ApJS, 57, 535
\bibitem[Coil et al.(2000)]{coi00} Coil, A. L., \& Ho, P. T. P., 2000. \apj, 533, 245
% Journals(e.g. A\&A,ApJ,AJ,NMRAS,PASP ...)
% \bibitem[Deguchi et al.(2004)]{deg04} Deguchi, S., et al. 2004, PASJ,  56, 261 %  Fujii, T.,  Glass, I. S., Imai, H.,  Ita, Y., Izumiura, H., Kameya, O., % Miyazaki, A., Nakada, Y., \& Nakashima, J.,  
\bibitem[Deguchi et al.(2002)]{deg02} Deguchi, S., Fujii, T.,  Miyoshi, M., \& Nakashima, J. 2002, PASJ, 54, 61
\bibitem[Deguchi et al.(2004)]{deg04} Deguchi, S., et al. 2004, \pasj,  56, 261
\bibitem[Figer et al.(2002)]{fig02} Figer, D.F., et al. 2002. ApJ 581, 258
\bibitem[Fujii et al.(2006)]{fuj06} Fujii, T., Deguchi, S., Ita, Y., Izumiura, H., Kameya, O., Miyazaki, A., \& Nakada, Y. 2006, PASJ 58, 529
\bibitem[Glass et al.~(2001)]{gla01} Glass, I. S., Matsumoto, S., Carter, B. S.,  \& Sekiguchi, K. 2001, MNRAS, 321, 77 
\bibitem[Genzel et al.(1990)]{gen90} Genzel, R., Stacey, G. J., Harris, A. I., Townes, C. H., Geis, N., 
  Graf, U. U., Poglitsch, A., \& Stutzki, J. 1990, \apj, 356, 160.
\bibitem[G\"{u}sten et al. (1981)]{gus81} G\"{u}sten, R., Walmsley, C.M., Pauls, T., 1981, A\&A 103, 197.
\bibitem[Handa et al.(2006)]{han06} Handa, T., Sakano, M., Naito, S., Hiramatsu, M., \& Tsuboi, M. 2006, \apj, 636, 261
\bibitem[Herbst et al.(1977)]{her77} Herbst, E., Green, S., Thaddeus, P., \& Klemperer, W. 1977, ApJ, 215, 503
\bibitem[Ho et al.(1985)]{ho85} Ho, P. T. P., Jackson, J. M., Barrett, A. H., \& Armstrong, J. T. 1985, \apj, 288, 575.
\bibitem[Ho et al.(1991)]{ho91} Ho, P. T. P., Ho, L. C., Szczepanski, J. C., Jackson, J. M.,
Armstrong, J. T., \& Barrett, A. H., 1991. Nature 350, 309.
% \bibitem[Imai et al.(2002)]{ima02} Imai, H., et al. 2002, PASJ,  54, L19 % Deguchi, S., Fujii, T.,  Glass, I. S., Ita, Y., Izumiura, H., Kameya, O., % Miyazaki, A., Nakada, Y., \& Nakashima, J.,  
\bibitem[Huettemeister et al.(1998)]{hue98} Huettemeister, S., Dahmen, G., Mauersberger, R., Henkel, C., Wilson, T. L. \&  Martin-Pintado, J. 1998, A\&A, 334, 646  % HOCO+ only in the fig5 
\bibitem[Imai et al.(2002)]{ima02} Imai, H., et al. 2002, \pasj,  54, L19
\bibitem[Izumiura et al.(1995)]{izu95} Izumiura, H., Deguchi, S., Hashimoto, O., Nakada, Y., Onaka, T., Ono, T., Ukita, N.,  \& Yamamura, I. 1995, ApJ,   453,  837 
% \bibitem[Jewell et al. (1991)]{jew91}Jewell, P.R., Snyder, L.E., Walmsley, C.M., Wilson, T.L.,  \& Gensheimer, P.D., 1991, \aap, 242, 211
%\bibitem[Justtanont et al. (1998)]{jus98} Justtanont, K., Feuchtgruber, H., de Jong, T., Cami, J., Waters, L. B. F. M., Yamamura, I., Onaka, T. 1998, \aap, 330, L17
\bibitem[Kaifu et al.(2004)]{kai04} Kaifu, N. et al. 
% Ohishi, Masatoshi; Kawaguchi, Kentarou; Saito, Shuji; Yamamoto, Satoshi; Miyaji, Takeshi; Miyazawa, Keisuke; Ishikawa, Shin-Ichi; Noumaru, Chiaki; Harasawa, Sumiko; and 2
 	    2004, PASJ, 56, 69	 
\bibitem[Lang et al.(2001)]{lan01} Lang, C. C.. Goss, W. M., \& Morris, M. 2001, \aj, 121, 2681
\bibitem[Lee et al.(2006)]{lee06} Lee, S. Pak, S., Choi, M., Davis, C. J., Geball, T. R.
 Herrnstein, R. M., Ho., P. T. P., Minh, Y. C., \& Lee, S. 2006 (submitted to ApJ)
\bibitem[Lilley \& Palmer(1968)]{lil68} Lilley, A. E. \& Palmer, P. 1968, \apjs, 16, 143
\bibitem[Lovas et al.(2004)]{lov04} Lovas, 2004, J. Phys. Chem. Ref. Data 33, 177
\bibitem[Martin-Pintado et al.(1997)]{mar97} Martin-Pintado, J., de Vicente, P., Fuente, A. \& Planesas, P. 1997, \apjl, 482, L45
\bibitem[Minh et al.(1998)]{min98} Minh, Y.C., Haikala, L., Hjalmarson, \AA., Irvine, W.M., 1998, \apj, 498, 261.
\bibitem[Minh et al.(2005)]{min05} Minh, Y. C., Kim, S.-J., Pak, Soojong, Lee, Sungho, Irvine, W. M., Nyman, L. 2005, New Astronomy, 10, 425
\bibitem[Minh et al.(1988)]{min88} Minh, Y. C., Irvine, W. M., \& Ziurys, L. M. 1988, \apj, 334, 175
\bibitem[Minh et al.(1991)]{min91} Minh, Y. C., Brewer, M. K., Irvine, W. M., Friberg, P., \& Johansson, L. E. B. 1991, A\&A, 244, 470
\bibitem[Miyazaki et al.(2001)]{miy01} Miyazaki, A., Deguchi, S., Tsuboi, M., Kasuga, T., \& Takano, S., 2001, PASJ, 53, 501
\bibitem[Miyazaki \& Tsuboi~(2000)]{miy00} Miyazaki, A., \& Tsuboi, M., 2000, \apj, 536, 357    
\bibitem[Murakami et al.(2003)]{mur03} Murakami, H., Senda, A., Maeda, Y., \& Koyama, K. 2003, Astro. Nachr., 234, 125
\bibitem[Oka et al.(1998)]{oka98} Oka, T., Hasegawa, T., Sato, F., Tsuboi, M., \& Miyazaki, A. 1998, \apjs, 118, 455 
\bibitem[Okumura et al.(1991)]{oku91} Okumura, S.K., et al. 
% Ishiguro, M., Fomalont, E.B., Hasegawa, T., Kasuga, T., 
% Morita, K.-I., Kawabe, R., Kobayashi, H., 
1991, ApJ 378, 127
% \bibitem[Pickett~(1991)]{pic91} Pickett, H. M. 1991, J. Molec. Spectroscopy 148, 371
\bibitem[Pickett et al.~(1998)]{pic98}  Pickett, H. M. , Poynter, R. L. , Cohen, E. A. , Delitsky,  M. L., Pearson, J. C., \& Muller, H. S. P. 
1998, J. Quant. Spectrosc. \& Rad. Transfer, 60, 883
\bibitem[Pineau des Forets, \& Flower~(1997)]{pin97} Pineau des Forets, G. and Flower, D. 1997, 
    IAU Symp. 178, Molecules in Astrophysics, ed. E. F. van Dishoeck, 113.
\bibitem[Serabyn et al.(1992)]{ser92} Serabyn, E., Lacy, J. H., \& Achterman, J. M., 1992. \apj, 395, 166.
\bibitem[Shiki et al.(1997)]{shi97} Shiki, S., Ohishi, M. \& Deguchi, S. 1997, \apj, 478, 206
\bibitem[Thaddeus et al.(1981)]{tha81} Thaddeus, P., Guelin, M., \& Linke, R. A. 1981, \apjl, 246, L41
\bibitem[Tuboi et al.(1999)]{tsu99} Tsuboi, M. Handa, T., \& Ukita, N., 1999, \apjs, 120, 1 
\bibitem[Turner et al.(1999)]{tur99} Turner, B.E., Terzieva, R., \& Herbst, E. 1999, \apj, 518, 699.
\bibitem[van Dishoeck~(2004)]{van04}van Dishoeck, E. F. 2004, ARA\&A, 42, 119
\bibitem[Yusef-Zadeh~(2003)]{yus03} Yusef-Zadeh, F. 2003, \apj, 598, 325.
\bibitem[Ziurys et al.(1989)]{ziu89} Ziurys, L. M., Friberg, P., \& Irvine, W. M. 1989, \apj, 343, 201

\end{thebibliography}
\end{document}